# Design of polarization-insensitive superconducting single photon detectors with high-index dielectrics

L. Redaelli[1,2],*, V. Zwiller[1,2,3,4], E. Monroy[1,2], J.M. Gérard[1,2]

[1] *Univ. Grenoble Alpes, F-38000 Grenoble, France*
[2] *CEA, INAC-PHELIQS, Nanophysics and Semiconductors group, F-38000, Grenoble, France*
[3] *TU Delft, Kavli Institute of Nanosciences, 2628 CH Delft, The Netherlands*
[4] *KTH Stockholm, Department of Applied Physics, 114 28 Stockholm, Sweden*
*\*luca.redaelli@cea.fr*

**Abstract:** In this paper, the design of superconducting-nanowire single-photon detectors which are insensitive to the polarization of the incident light is investigated. By using high-refractive-index dielectrics, the index mismatch between the nanowire and the surrounding media is reduced. This enhances the absorption of light with electric field vector perpendicular to the nanowire segments, which is generally hindered in this kind of detectors. Building on this principle and focusing on NbTiN nanowire devices, we present several easy-to-realize cavity architectures which allow high absorption efficiency (in excess of 90%) and polarization insensitivity simultaneously. Designs based on ultranarrow nanowires, for which the polarization sensitivity is much more marked, are also presented. Finally, we briefly discuss the specific advantages of this approach in the case of WSi or MoSi nanowires.

1. **Introduction**

Superconducting-nanowire single photon detectors (SNSPDs) find application today in a wide range of fields [1], from quantum optics experiments [1-3] to the characterization of single-photon emitters [5], from satellite communication [6] to life sciences [7]. In recent years, remarkable progress has been made in boosting the detection efficiency of SNSPDs operating in the near infrared. Using WSi nanowires, a record efficiency of 93% has been attained [8]. More recently, efficiencies well in excess of 80% have been demonstrated with NbTiN [9] and MoSi [10] nanowires.

All these devices make use of a superconducting film, patterned as a meandering nanowire, and enclosed in a resonant optical cavity to enhance absorption. In this configuration, the meandering nanowire acts as a sub-wavelength grating and interacts in different ways with the incoming light according to its polarization. Transverse-electric (TE) polarized light, where the electric field oscillates parallel to the longer nanowire segments, is absorbed much more efficiently than transverse-magnetic (TM) polarized light, where the electric field oscillates perpendicularly to the longer nanowire segments [11]. It is possible to partly counterbalance this effect by designing wider nanowires, but at the price of degrading the internal efficiency. In practical devices operating in the near infrared, the width rarely exceeds 100 nm for NbTiN, and 120-130 nm for WSi or MoSi nanowires. In the mid-infrared, a spectral region where many potential applications are emerging [12, 13], even narrower nanowires are needed so as to cope with a reduced photon energy, and thus TM absorption becomes even weaker.

In order to achieve polarization insensitivity, it has been recently proposed to embed the superconducting nanowire within a thin (~25 nm) film of high permittivity material, such as silicon [14]. In this reference, the so-embedded nanowire is placed in a cavity to boost both

TE and TM absorption efficiencies. Although polarization-insensitivity over a wide wavelength range is actually predicted, such a design would be very difficult to implement in practice: after embedding the nanowire in the thin Si film, the surface would need to be perfectly planarized, in order to realize a 54-nm-wide Si nanowire grid on top. Furthermore, the use of silicon prevents the use of this design for detectors operating below the wavelength of 1.1 µm.

In this paper, we introduce novel microcavity designs which provide both high absorption efficiency and TE-TM polarization insensitivity. We show that the insertion of a simple layer of high-index (n > 2) dielectric material, either placed below or above the superconducting nanowire, enables to compensate the TE-TM absorption efficiency difference to a very large extent, and is fully compatible with the use of a microcavity to maximize both absorption coefficients simultaneously. This approach leads to simple designs, rather close to those already in use from a fabrication point of view, which should greatly facilitate their practical implementation in commercial devices.

After introducing the optical model in section 2, we briefly discuss in section 3 the physical reasons that make high-refractive-index dielectrics a key resource for polarization insensitive SSPDs, and we analyze by finite-difference time-domain (FDTD) simulations the effect of the surrounding media indices on the absorption efficiency of the nanowire. In section 4, we apply these basic principles to the simplest cavity architectures, and propose an innovative, easy-to-realize design which offers polarization-insensitive absorption efficiency in excess of 90% using an NbTiN nanowire. Finally, in section 5 we present a cavity design where a narrower NbTiN nanowire is used, and briefly discuss the specific advantages of this approach in the case of WSi or MoSi nanowires.

**2. Optical model and simulation method**

In Fig. 1 the simplest simulated structure is schematically represented. It consists of a 7 nm-thick meandering NbTiN nanowire sandwiched between a substrate and a superstrate. The nanowire segments are 100 nm wide, the gap between two adjacent nanowire segments ("meander gap") is 100 nm as well. The "gap filling", i.e. the dielectric material filling the meander gap, is the same as the superstrate material.

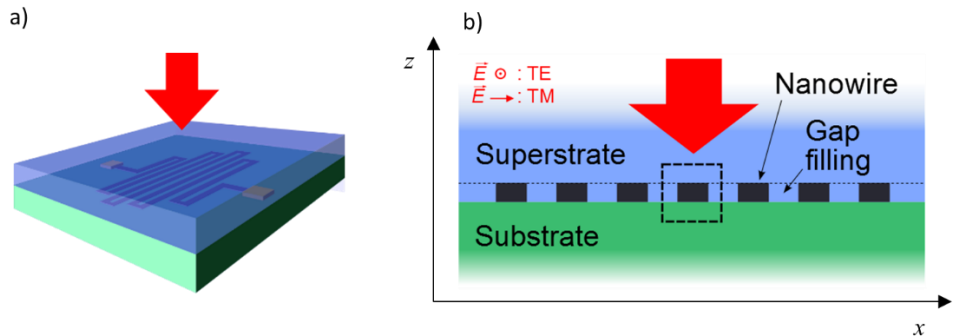

Fig. 1: (a) Illustration of the meandering-nanowire SNSPD studied in this paper. The red arrow marks the direction of the incident photons, which we always consider as incident through the superstrate. (b) Cross-section of (a). The simulated domain roughly corresponds to the area surrounded by the dashed square line. The drawings are not to scale.

For the purpose of calculation, the device is modeled as an infinite grating, as shown in Fig. 1(b), and the incident light is approximated by a plane wave. FDTD calculations have been performed using the commercial software *RSoft FullWAVE* [15]. The simulation considers a single grating period with in-plane periodic boundary conditions. The top and bottom domain boundaries are 0.2 µm-thick perfectly matched layers (PML). A non-uniform graded grid is used: the grid is finest inside the thin nanowires (0.7 nm in z) and close to the layer boundaries (1 nm in x, 0.7 nm in z). The coarser grid size is 10 nm, both in x and z. The absorption efficiency is calculated, once the steady state is reached, by subtracting from the input power the power reaching the top PML boundary due to reflection, and the transmitted power reaching the bottom PML boundary (or the power absorbed by the backside mirror, if any).

The wavelength-dependent complex refractive indices used in the calculations were taken from literature. Their values at 1.55 µm are: 4.80 + $i$6.05 for NbTiN [16], 4.58 + $i$3.66 for WSi [8], 2.25 for $TiO_2$ [8], 1.44 for $SiO_2$ [15], 3.47 for Si [15], and 0.57 + $i$9.66 for Au [15].

## 3. Polarization sensitivity and surrounding media indices

The absorption of an optically thin metallic film is not only a function of the film complex refractive index, but also of the substrate and of the incident medium ("superstrate") indices [17]. In the case of a subwavelength grating, the absorption can be modeled using the effective medium theory (EMT) [18,19], which approximates the grating by a uniform birefringent film with different refractive indices for TE and TM radiation. However, if we consider a grating of metallic wires, EMT works for TE-polarized light, but fails for the TM polarization [17,20].

When excited by TM-polarized light, the electric field distribution in the nanowire is highly inhomogeneous. If the gap between two wire segments is filled by air, there is a field minimum in the nanowire, close to the boundaries, and a gradual increase of the field intensity towards the nanowire center. The cause of this gradient is the permittivity mismatch between the surrounding medium (air) and the nanowire. Maxwell's equations impose that the normal component of the electric displacement vector D is continuous at the interfaces: since D is the product of the material permittivity constant ε and the electric field E, a mismatch in ε leads to a mismatch in E. It follows that, to enhance the electric field in the nanowire, i.e. to enhance TM absorption, the permittivity mismatch between the surrounding dielectric and the nanowire should be reduced [14].

In Figs. 2(a) and 2(b) the nanowire absorption efficiency is calculated as a function of the substrate and the superstrate indices, for TE and TM radiation respectively. The behavior is qualitatively the same for both polarizations, except when both superstrate and substrate indices are low. This feature is highlighted in the polarization-sensitivity plot of Fig. 2(c). The polarization sensitivity is defined as: $S = (\eta_{abs,TE} - \eta_{abs,TM}) / (\eta_{abs,TE} + \eta_{abs,TM})$, where $\eta_{abs,TE}$ and $\eta_{abs,TM}$ are the absorption efficiencies for TE and TM polarized light, respectively. For superstrate and substrate indices close to 1, $S$ is maximum (high TE absorption, low TM absorption). Increasing one or both indices, on the other hand, reduces the polarization sensitivity, which can be close to zero when both indices exceed 2.5.

As pointed out earlier, in these calculations the gap filling index is the same as the superstrate index, as it is the case in real front-injection SNSPDs. If the calculations are repeated by choosing the gap filling index to be the same as the "substrate" index (this is the case when, in real devices, light is injected through the carrier wafer), the trends are roughly the same, as shown in Fig. 2(d): the color map, illustrating the dependence of sensitivity over substrate and superstrate indices, is very similar to the one in (c), only with the x and y axes inverted. It is important to keep in mind that the influence of the gap filling index on TE absorption is negligible, and it is the TM absorption that varies, being favored by higher gap filling indices. For the considered index range the difference in TM absorption can be as large as 20%.

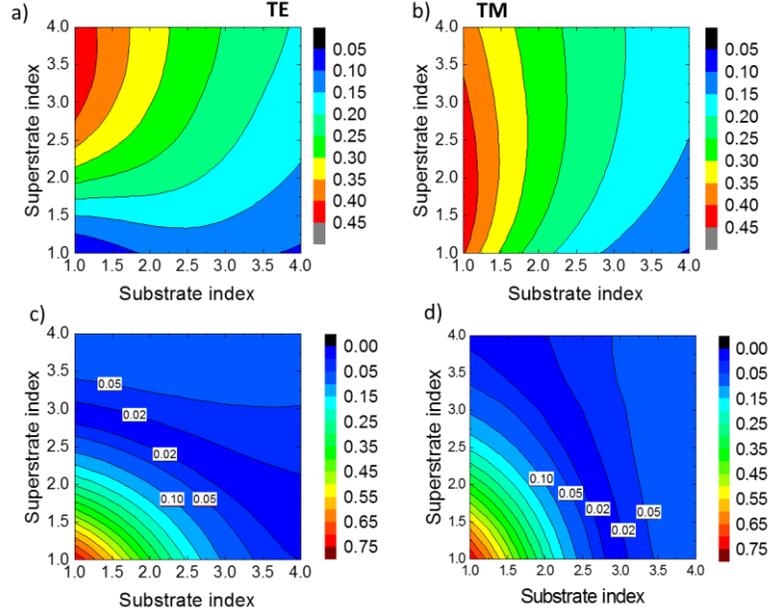

Fig. 2: (a) Absorption efficiency for TE polarized light and (b) TM polarized light as a function of substrate and superstrate index. The indices are varied in steps of 0.2 between 1 and 4, and the results interpolated to form a continuous 2D color map. (c) Polarization sensitivity calculated from (a) and (b). (d) Polarization sensitivity, calculated in the case where the gap filling index is equal to the substrate index. (As compared to (a), (b) and (c), where the gap filling index is equal to the superstrate index)

According to the calculations in Fig. 2, the maximum possible polarization-independent absorption efficiency of a 7 nm-thick NbTiN wire, in absence of cavity, is achieved for a substrate index of 1 and a superstrate index of about 3. The simplest practical implementation of this design would be a nanowire fabricated on a Si substrate (index 3.47), with backside light injection and backside antireflection coating. Taking into account that, in such a case, the gap filling index would be 1 as well, we obtain an absorption efficiency of 38.3% for TE, 38.9% for TM.

## 4. Polarization-insensitivity in simple cavity structures

In order to enhance the absorption efficiency, different cavity designs can be found in literature. The simplest cavity consists of a backside reflector, such as a metallic mirror, and of a quarter-wave "spacer", usually made of silicon dioxide [9, 11, 21-26]. We study here the impact of the spacer and superstrate indices on the nanowire absorption efficiency. We consider a nearly perfect reflector as backside mirror (index $0 + i1000$), and a spacer thickness of exactly $\lambda/4n_{spa}$, where $\lambda$ is the vacuum wavelength and $n_{spa}$ the spacer index. The calculations results, shown in Figs. 3(b) and 3(c), indicate that the spacer index has no impact on TE absorption, whereas high-index materials should be preferred to enhance TM absorption.

It is important to note that the optimum optical thickness of the spacer is different for TE and for TM, since the cavity resonance is modified by the birefringent behavior of the grating. Moreover, the optimum optical thickness of the spacer changes as a function of the spacer index (slightly for TE, but more significantly for TM) as shown in Figs. 3(d) and 3(e). The optima are farther apart when the index is low, but they get closer to $\lambda/4n_{spa}$ and to each other for higher index.

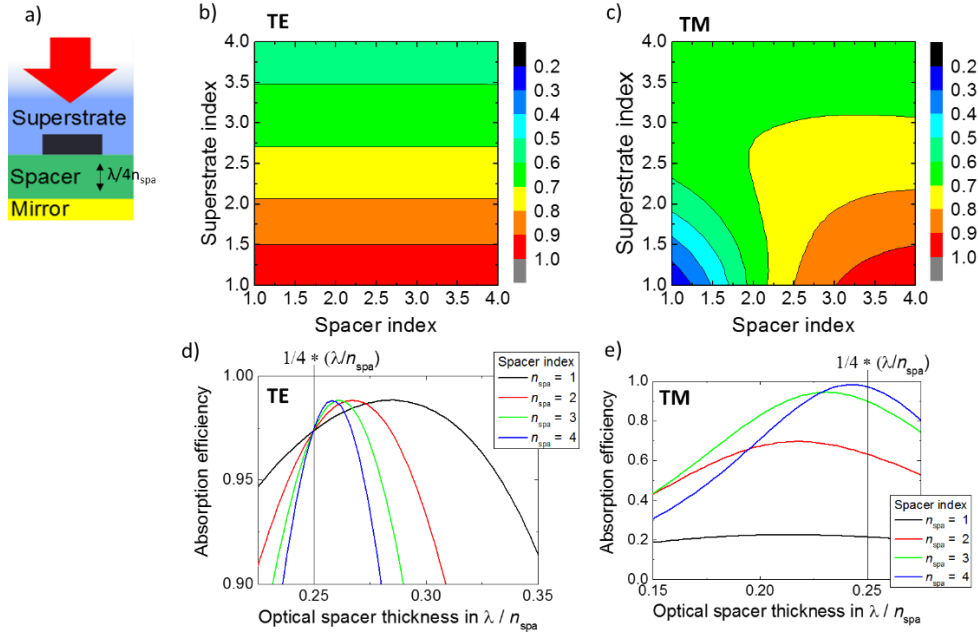

Fig. 3: (a) Simple cavity design. (b) Absorption efficiency for TE polarized light and (c) TM polarized light as a function of spacer and superstrate indices, for the cavity design shown in (a). (d) TE and (e) TM absorption efficiency as a function of the spacer thickness and spacer index. Note that the vertical scale is different in the two plots.

According to these calculations, high-index dielectrics should be preferred for the fabrication of polarization-independent SNSPDs. Similar conclusions were drawn in ref. [27], where a design was proposed in which the nanowire would be directly patterned on top of a Si spacer, and a gold mirror would be deposited on the backside after deep silicon etching. The proposed realization of this structure is however challenging, since the use of silicon-on-insulator (SOI) is needed to provide an etch stop. Furthermore, as pointed out previously, the use of Si precludes the use of this architecture for wavelengths below 1.1 µm.

In Fig. 4(a), we propose a new design which can achieve similar performance, but is much simpler to implement. A high-index front layer ("matching layer") is deposited on top of the traditional SiO$_2$/Au cavity, in order to reduce the index mismatch and allow highly efficient TM absorption. The effect of the matching layer on TE absorption is minimized by setting its thickness to $\lambda/2n_{ML}$, where $n_{ML}$ is the refractive index of the dielectric material used in the matching layer. Figures 4(b) and (c) present the calculated nanowire absorption efficiency for TE and TM polarized light, respectively, as a function of the refractive indices of the spacer and matching layer. It is important to highlight that Figs. 4(b) and 4(c) are calculated for a constant optical spacer thickness of exactly $\lambda/4n_{spa}$. Once a material is chosen for the front matching layer, it is possible to further tune the absorption by adjusting the spacer thickness, as shown in Fig. 4(d), until polarization-insensitivity is achieved.

The simplest realization of this cavity architecture makes use of a gold mirror, a SiO$_2$ spacer, and a 344 nm-thick TiO$_2$ front matching layer. By tuning the SiO$_2$ spacer thickness to 216 nm, it is possible to achieve a polarization-insensitive absorption efficiency of 90.8% at the target wavelength of 1.55 µm, as shown in Figs. 4(d) and 4(e). Even higher polarization-independent absorption efficiencies can be achieved by using higher-index dielectrics. For example, the SiO$_2$ spacer could be easily replaced by TiO$_2$, so that the nanowire is fully embedded in TiO$_2$, boosting the polarization-insensitive absorption efficiency to 92.7%. Note that the TiO$_2$ refractive index is considered in this paper to be 2.25, but it can be increased above 2.4 by adjusting the material deposition parameters [28]. This would lead to a further increase of the absorption efficiency.

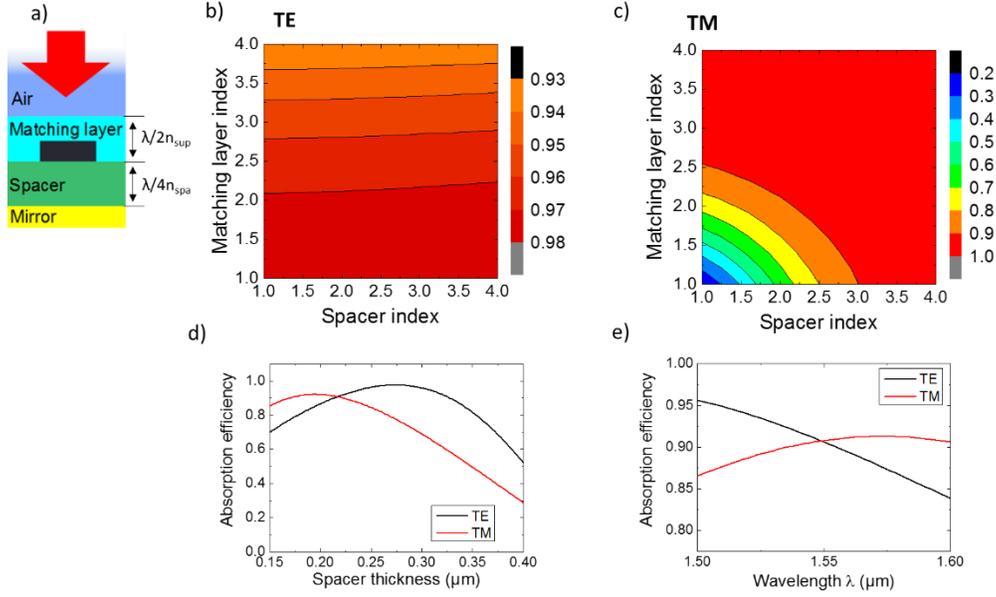

Fig. 4: (a) Cavity design making use of a $\lambda/2n_{ML}$ matching layer. (b) Absorption efficiency for TE polarized light and (c) TM polarized light as a function of spacer and matching layer index, for the cavity design shown in (a). (d) TM absorption efficiency as a function of the SiO$_2$ spacer thickness for a Au/SiO$_2$/TiO$_2$ cavity, showing the crossing point at 216 nm. (e) Absorption efficiency as a function of wavelength in the optimized structure.

## 5. Narrow nanowires and low-absorption superconducting materials

As pointed out earlier, there is great interest in efficient mid-infrared detectors based on narrower superconducting nanowires. However, the TM absorption efficiency quickly drops when decreasing the nanowire width. Furthermore, to cover the same spot area, either much longer nanowires are needed, or the fill factor needs to be reduced, which would further decrease the absorption efficiency.

By combining high–index dielectrics and a stronger cavity, it is possible to realize polarization-insensitive narrow-nanowire detectors with high detection efficiency, even if the fill factor is reduced. Let us consider for example the cavity design in Fig. 5(a). The nanowire width is reduced down to 50 nm whereas the gap is kept at 100 nm, so that the fill factor is decreased from 50% to 33%. Decreasing the fill factor to 33% or below has the further advantage that the radius of the 180° bends of the meandering nanowire can be optimized to avoid current crowding at the corners [29], which has a positive impact on the device performance. In this design, the nanowire is embedded in a high-index cavity made of TiO$_2$. Since the nanowire absorption is weaker than in the previous examples due to the lower fill factor, a front mirror, made of one SiO$_2$/TiO$_2$ pair, is added to the front-side. The backside mirror is made of Au. As shown by the calculations in Fig. 5(b), by optimizing the layer thicknesses the peak polarization-independent absorption efficiency can reach 84.1%.

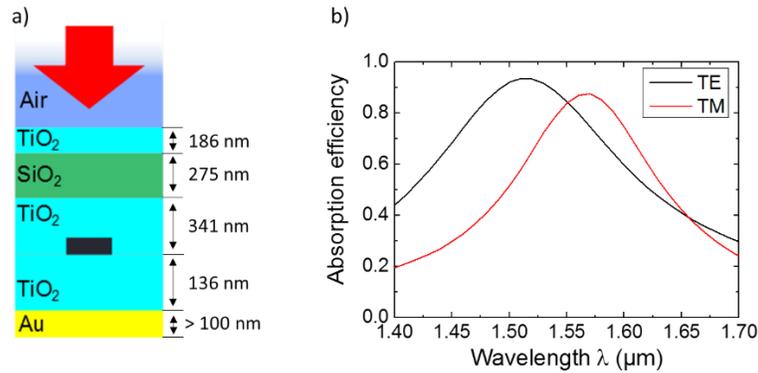

Fig. 5: (a) Cavity design optimized for polarization-insensitive absorption efficiency of a 50 nm-wide NbTiN nanowire. (b) Absorption efficiency as a function of wavelength in the optimized structure.

An important loss mechanism in this architecture is the absorption of radiation by the Au mirror, which is as high as 4.1% for TE and 8.1% for TM at 1.55 µm. The detection efficiency can be slightly improved, up to 85.7%, by using an Ag mirror instead. The use of a $TiO_2/SiO_2$ Bragg mirror is not advisable, since the TE and TM absorption peaks would become spectrally narrower, and hence the polarization-independent absorption would be less efficient. Further improvements of this structure would be possible, as for the one presented in Section 2, by using higher-index dielectrics.

Finally, it is interesting to consider the case of WSi and MoSi nanowires, which are reported to show lower polarization dependence that NbN or NbTiN. This is partly due to the fact that, in the reported devices [8,10], wider and thinner MoSi and WSi nanowires have been implemented. In addition, the imaginary part of the refractive index of MoSi and WSi is lower than that of NbN/NbTiN, which further reduces the polarization sensitivity at the price of a decrease of the TE absorption efficiency for the same nanowire dimensions. Nonetheless, the basic principles presented in this paper are still valid for devices based on WSi and MoSi nanowires. Higher-index dielectrics in proximity of the nanowire enhance TM absorption and cause a spectral shift of the TE and TM absorption peaks that brings them closer to each other. However, a stronger microcavity is needed to bring the efficiency close to unity.

As an illustration, let us consider for example the cavity structures in Figs. 6(a) and 6(c), which use the same materials and the same nominal nanowire geometry as reference [8] (4.5-nm-thick, 120-nm-wide WSi nanowires, 80 nm meander gap). The design in (a) makes use of a 10-period Bragg backside mirror and a two-period front dielectric mirror, tuned to achieve the highest possible polarization-insensitive absorption. The design in Fig. 6(c) is conceptually identical, with the addition of two high-index inner cavity layers on both sides of the nanowire. The results in Figs. 5(b) and 5(d) show that in the first case the absorption maxima are further apart, and the maximum achievable polarization-insensitive absorption efficiency is slightly below 98%. For the cavity making use of $TiO_2$ inner layers, on the other hand, a polarization-insensitive absorption efficiency of about 99.5% could be reached.

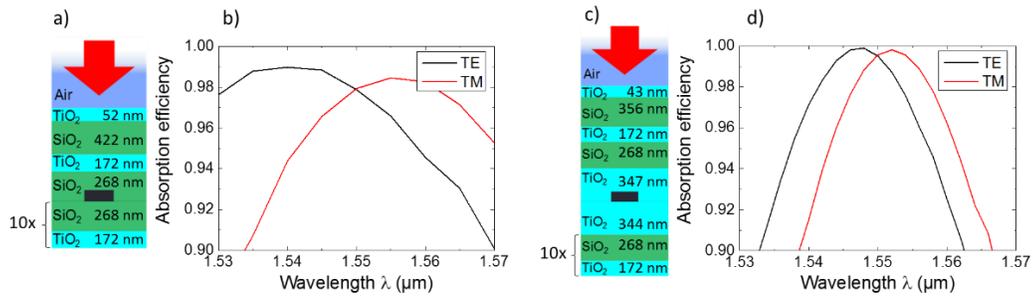

Fig. 6: (a) and (c): cavity designs for near-unity, polarization independent absorption efficiency making use of WSi nanowire, and respective simulated performance [(b) and (d)].

## 6. Summary and conclusions

In this paper, the interest of using high-index materials for the realization of SNSPDs has been shown. High-index dielectric layers in proximity of the superconducting nanowire reduce the index mismatch between the nanowire and the surrounding medium, considerably enhancing the absorption efficiency of TM polarized light. In a detector consisting of a 7 nm-thick NbTiN nanowire, absorption efficiencies approaching 40% for both TE and TM polarized light can be achieved by simply fabricating the nanowire on bare silicon (index 3.47).

By enclosing the nanowire in an optical cavity, TE absorption efficiencies approaching unity can be easily achieved. High-index dielectrics can be used, in this case, to boost TM absorption and realize highly efficient polarization-insensitive detectors. An innovative design is proposed in this paper, where a high-index front layer is added to a simple $SiO_2$/Au cavity. This device would achieve polarization-insensitive absorption efficiency in excess of 90%.

High-index dielectrics are especially useful when dealing with ultranarrow nanowires (well below 100 nm in width), where TM absorption is particularly inefficient. A design making use of a 50 nm-wide meandering nanowire has been shown, where polarization-insensitive absorption efficiency of about 85% can be achieved, thanks to the high-index cavity, in spite of a low fill factor (33%).

Finally, polarization-sensitivity is usually less marked in WSi or MoSi-based detectors, since they typically make use of wider and thinner nanowires, but also because the absorption coefficient of these materials is lower than that of NbTiN. The advantages of using high index dielectrics are therefore less marked for this kind of SNSPDs. Nevertheless, by using high-index dielectrics in proximity of the nanowire, it is possible to achieve absorption efficiencies which go beyond what is theoretically possible with low-index dielectrics (such as $SiO_2$). A design making use of a wide WSi nanowire and $TiO_2$ inner cavity layers has been shown, which can achieve polarization-insensitive absorption efficiency of 99.5%, while the "classical" design based on SiO2 is limited by theory below 98%.

## 7. Funding

This work was funded by The European Commission via the *Marie Skłodowska Curie IF grant* "SuSiPOD" (H2020-MSCA-IF-2015, #657497), the *French National Research Agency* via the "WASI" (ANR-14-CE26-0007) program, and the Grenoble *Nanosciences Foundation*.

## References

1. C. M. Natarajan and M. G. Tanner, "Superconducting nanowire single-photon detectors: physics and applications," Supercond. Sci. Technol. **25**, 063001 (2012).
2. P. V. Klimov, A. L. Falk, D. J. Christle, V. V. Dobrovitski, and D. D. Awschalom, "Quantum entanglement at ambient conditions in a macroscopic solid-state spin ensemble," Sci. Adv. **1**, e1501015 (2015).
3. A. Lenhard, J. Brito, S. Kucera, M. Bock, J. Eschner, and C. Becher, "Single telecom photon heralding by wavelength multiplexing in an optical fiber," Appl. Phys. B **122**, (2016).
4. C. Xiong, X. Zhang, Z. Liu, M. J. Collins, A. Mahendra, L. G. Helt, M. J. Steel, D.-Y. Choi, C. J. Chae, P. H. W. Leong, and B. J. Eggleton, "Active temporal multiplexing of indistinguishable heralded single photons," Nat. Commun. **7**, 10853 (2016).
5. X. Ma, N. F. Hartmann, J. K. S. Baldwin, S. K. Doorn, and H. Htoon, "Room-temperature single-photon generation from solitary dopants of carbon nanotubes," Nat. Nanotechnol. **10**, 671–675 (2015).
6. M. Grein, E. Dauler, A. Kerman, M. Willis, B. Romkey, B. Robinson, D. Murphy, and D. Boroson, "A superconducting photon-counting receiver for optical communication from the Moon," SPIE Newsroom, 9 July 2015.
7. N. R. Gemmell, A. McCarthy, B. Liu, M. G. Tanner, S. D. Dorenbos, V. Zwiller, M. S. Patterson, G. S. Buller, B. C. Wilson, and R. H. Hadfield, "Singlet oxygen luminescence detection with a fiber-coupled superconducting nanowire single-photon detector," Opt. Express **21**, 5005 (2013).
8. F. Marsili, V. B. Verma, J. A. Stern, S. Harrington, A. E. Lita, T. Gerrits, I. Vayshenker, B. Baek, M. D. Shaw, R. P. Mirin, and S. W. Nam, "Detecting single infrared photons with 93% system efficiency," Nat. Photonics 7, 210–214 (2013).
9. L. Redaelli, G. Bulgarini, S. Dobrovolskiy, S. N. Dorenbos, V. Zwiller, E. Monroy, and J. M. Gérard, "Design of broadband high-efficiency superconducting-nanowire single photon detectors," Supercond. Sci. Technol. **29**, 065016 (2016).
10. V. B. Verma, B. Korzh, F. Bussières, R. D. Horansky, S. D. Dyer, A. E. Lita, I. Vayshenker, F. Marsili, M. D. Shaw, H. Zbinden, R. P. Mirin, and S. W. Nam, "High-efficiency superconducting nanowire single-photon detectors fabricated from MoSi thin-films," Opt. Express **23**, 33792 (2015).
11. V. Anant, A. J. Kerman, E. A. Dauler, J. K. W. Yang, K. M. Rosfjord, and K. K. Berggren, "Optical properties of superconducting nanowire single-photon detectors," Opt. Express **16**, 10750 (2008).
12. F. Marsili, F. Bellei, F. Najafi, A. E. Dane, E. A. Dauler, R. J. Molnar, and K. K. Berggren, "Efficient Single Photon Detection from 500 nm to 5 μm Wavelength," Nano Lett. **12**, 4799–4804 (2012).
13. F. Bellei, A. P. Cartwright, A. N. McCaughan, A. E. Dane, F. Najafi, Q. Zhao, and K. K. Berggren, "Free-space-coupled superconducting nanowire single-photon detectors for infrared optical communications," Opt. Express **24**, 3248 (2016).
14. F. Zheng, R. Xu, G. Zhu, B. Jin, L. Kang, W. Xu, J. Chen, and P. Wu, "Design of a polarization-insensitive superconducting nanowire single photon detector with high detection efficiency," Sci. Rep. **6**, 22710 (2016).
15. RSoft Photonic Design software, Synopsys® (http://optics.synopsys.com/rsoft/)
16. S. N. Dorenbos, "Superconducting Single Photon Detectors," PhD Dissertation, Technische Universiteit Delft (2011).
17. E. F. C. Driessen, F. R. Braakman, E. M. Reiger, S. N. Dorenbos, V. Zwiller, and M. J. A. de Dood, "Impedance model for the polarization-dependent optical absorption of superconducting single-photon detectors," Eur. Phys. J. – Appl. Phys. **47**, 10701 (2009).
18. A. Yariv and P. Yeh, "Electromagnetic propagation in periodic stratified media. II Birefringence, phase matching, and x-ray lasers" J. Opt. Soc. Am. **67**, 438 (1977).
19. P. Yeh, "A new optical model for wire grid polarizers," Opt. Commun. **26**, 289–292 (1978).
20. S. Moon and D. Kim, "Fitting-based determination of an effective medium of a metallic periodic structure and application to photonic crystals," J. Opt. Soc. Am. A **23**, 199 (2006).
21. K. M. Rosfjord, J. K. W. Yang, E. A. Dauler, A. J. Kerman, V. Anant, B. M. Voronov, G. N. Gol'tsman, and K. K. Berggren, "Nanowire single-photon detector with an integrated optical cavity and anti-reflection coating," Opt. Express **14**, 527 (2006).
22. B. Baek, J. A. Stern, and S. W. Nam, "Superconducting nanowire single-photon detector in an optical cavity for front-side illumination," Appl. Phys. Lett. **95**, 191110 (2009).
23. S. Miki, M. Takeda, M. Fujiwara, M. Sasaki, and Z. Wang, "Compactly packaged superconducting nanowire single-photon detector with an optical cavity for multichannel system," Opt. Express **17**, 23557 (2009).
24. M. G. Tanner, C. M. Natarajan, V. K. Pottapenjara, J. A. O'Connor, R. J. Warburton, R. H. Hadfield, B. Baek, S. Nam, S. N. Dorenbos, E. B. Ureña, T. Zijlstra, T. M. Klapwijk, and V. Zwiller, "Enhanced telecom wavelength single-photon detection with NbTiN superconducting nanowires on oxidized silicon," Appl. Phys. Lett. **96**, 221109 (2010).

25. D. Rosenberg, A. J. Kerman, R. J. Molnar, and E. A. Dauler, "High-speed and high-efficiency superconducting nanowire single photon detector array," Opt. Express **21**, 1440–1447 (2013).
26. S. Miki, T. Yamashita, H. Terai, and Z. Wang, "High performance fiber-coupled NbTiN superconducting nanowire single photon detectors with Gifford-McMahon cryocooler," Opt. Express **21**, 10208–10214 (2013).
27. H.-Y. Yin, H. Cai, R.-S. Cheng, Z. Xu, Z.-N. Jiang, J.-S. Liu, T.-F. Li, and W. Chen, "Polarization independent superconducting nanowire detector with high-detection efficiency," Rare Metals **34**, 71–76 (2015).
28. J. Kischkat, S. Peters, B. Gruska, M. Semtsiv, M. Chashnikova, M. Klinkmüller, O. Fedosenko, S. Machulik, A. Aleksandrova, G. Monastyrskyi, Y. Flores, and W. Ted Masselink, "Mid-infrared optical properties of thin films of aluminum oxide, titanium dioxide, silicon dioxide, aluminum nitride, and silicon nitride," Appl. Optics **51**, 6789 (2012).
29. A. Engel, J. J. Renema, K. Il'in, and A. Semenov, "Detection mechanism of superconducting nanowire single-photon detectors," Supercond. Sci. and Technol. **28**, 114003 (2015).